\documentclass[fleqn,10pt]{wlscirep}
\usepackage[utf8]{inputenc}
\usepackage[T1]{fontenc}
\usepackage{hyperref}
\usepackage{url}
\usepackage{times}
\usepackage{epsfig}
\usepackage{graphicx}
\usepackage{amsmath}
\usepackage{amssymb}
\usepackage{mathtools, nccmath}
\usepackage{svg}

\usepackage{booktabs}  

\usepackage{bm}
\usepackage{microtype}
\usepackage{amsmath}
\usepackage{amssymb}
\usepackage{amsthm}
\usepackage{array}
\usepackage{multirow}
\usepackage{cases}
\usepackage{subcaption}
\usepackage{algorithm}
\usepackage{algpseudocode}
\newtheorem{clm}{Claim}
\newtheorem{defi}{Definition}

\theoremstyle{remark}

\usepackage{txfonts}
\usepackage{xcolor}

\usepackage{pifont}
\newcommand{\cmark}{\ding{51}}%
\newcommand{\xmark}{\ding{55}}%
\def\ie{\emph{i.e}., }
\def\eg{\emph{e.g}., }

\def\etal{\emph{et al}. }
\def\wrt{\emph{w.r.t}. }

\title{Automated causal inference in application to randomized controlled clinical trials}

\author[1,*]{Ji~Q.~Wu}
\author[2]{Nanda~Horeweg}
\author[3]{Marco~de~Bruyn}
\author[2,**]{Remi~A.~Nout}
\author[4]{Ina~M.~J{\"u}rgenliemk-Schulz}
\author[5]{Ludy~C.H.W.~Lutgens}
\author[6,***]{Jan~J.~Jobsen}
\author[7]{Elzbieta~M.~van~der~Steen-Banasik}
\author[3]{Hans~W.~Nijman}
\author[8]{Vincent~T.H.B.M.~Smit}
\author[8]{Tjalling~Bosse}
\author[2]{Carien~L.~Creutzberg}
\author[1,*]{Viktor~H.~Koelzer}

\affil[1]{Department of Pathology and Molecular Pathology, University Hospital, University of Zurich, Zurich, Switzerland.}
\affil[2]{Department of Radiation Oncology, Leiden University Medical Center, Leiden, The Netherlands.}
\affil[3]{Department of Obstetrics and Gynecology, University of Groningen, University Medical Center Groningen, Groningen, The Netherlands.}
\affil[4]{Department of Radiation Oncology, University Medical Center Utrecht, Utrecht, The Netherlands.}
\affil[5]{Maastricht Radiation Oncology Clinic, Maastricht, The Netherlands.}
\affil[6]{Department of Radiotherapy, Medisch Spectrum Twente, Enschede, The Netherlands.} 
\affil[7]{Radiotherapiegroep, Arnhem, The Netherlands.}
\affil[8]{Department of Pathology, Leiden University Medical Center, Leiden, The Netherlands.}
\affil[*]{Corresponding authors, Jiqing.Wu@usz.ch, Viktor.Koelzer@usz.ch}
\affil[**]{Currently employed at department of Radiotherapy, Erasmus MC Cancer Institute, University Medical Center Rotterdam, Rotterdam, The Netherlands.}
\affil[***]{Currently employed at department of Clinical Epidemiology, Medisch Spectrum Twente, Enschede, The Netherlands.}

\begin{abstract}
Randomized controlled trials (RCTs) are considered as the gold standard for testing causal hypotheses in the clinical domain. However, the investigation of prognostic variables of patient outcome in a hypothesized cause-effect route is not feasible using standard statistical methods. Here, we propose a new automated causal inference method (AutoCI) built upon the invariant causal prediction (ICP) framework for the causal re-interpretation of clinical trial data. Compared to existing methods, we show that the proposed AutoCI allows to efficiently determine the causal variables with a clear differentiation on two real-world RCTs of endometrial cancer patients with mature outcome and extensive clinicopathological and molecular data. This is achieved via suppressing the causal probability of non-causal variables by a wide margin. In ablation studies, we further demonstrate that the assignment of causal probabilities by AutoCI remain consistent in the presence of confounders. In conclusion, these results confirm the robustness and feasibility of AutoCI for future applications in real-world clinical analysis.
\end{abstract}
\begin{document}

\flushbottom
\maketitle
%
%
\thispagestyle{empty}

\section*{Introduction}
Many clinical studies are driven by the research questions that are statistical at first glance but causal by nature \cite{pearl2001causal}. For instance, how safe and efficient is a vaccine against viral infection \cite{voysey2021safety}? How are clinicopathological variables related to cancer patient survival \cite{horeweg2020prognostic}? From the causal perspective, the common ground of these problems starts with determining the causal variables of the outcome of interest.
In clinical medicine, randomized controlled trials (RCTs) are considered to be the gold standard to investigate cause-effect relationships \cite{hariton2018randomised}. In a prototypical RCT, a participant is randomly assigned to the experimental or control arm and the outcome of interest is observed. In the context of causal inference, such a randomization can be modeled with do-intervention \cite{peters2017elements}. 

In this article, we develop a new automated causal inference method (AutoCI) and apply this to two large-scale, practice changing RCTs of endometrial carcinoma patients conducted in the Netherlands from 1990-1997 (PORTEC 1 \cite{creutzberg2000surgery,creutzberg2011fifteen}) and 2002-2006 (PORTEC 2 \cite{nout2010vaginal,wortman2018ten}),  with full clinicopathological datasets and mature outcome data.
Endometrial carcinoma is the most common type of gynaecological cancer for women in developed countries \cite{sung2021global}. 
The majority of women with endometrial cancer (EC) that are diagnosed with early-stage disease, have a favorable prognosis and are treated with surgery \cite{van2021adjuvant}. Approximately 15–20\% of patients have an unfavorable prognosis with a high risk of distant metastasis \cite{van2021adjuvant}. For those patients, different adjuvant therapies, such as vaginal brachytherapy, external beam radiotherapy and chemoradiation, are recommended based on their risk group \cite{concin2021esgo}. The two trials (PORTEC 1 and 2) used in this study made a key contribution to clinical practice by investigating how these therapies impact the risk of recurrence rates and survival \cite{creutzberg2000surgery,nout2010vaginal}. 
According to the latest ESGO/ESTRO/ESP guidelines \cite{concin2021esgo} for the management of patients with endometrial carcinoma, the risk classification is based on a series of clinical and pathological variables such as tumor grading (Grade), lymphovascular space invasion (LVSI), myometrial invasion, etc. and molecular variables including but not limited to polymerase epsilon mutant EC (POLEmut), mismatch repair deficient (MMRd) EC, p53 abnormal EC (p53abn) and EC with no specific molecular profile (NSMP) \cite{concin2021esgo}. In a recent study \cite{horeweg2020prognostic}, correlative statistical methods were used to investigate the hazardous relevance of these variables to EC recurrence. There is strong evidence to suggest that these variables impact EC recurrence, but a systematic investigation from a modern causal inference perspective to support this understanding has not been done.

Causal inference addresses the determination of cause and effect relationships from data \cite{hernan2020causal,zenil2019causal,prosperi2020causal,luo2020causal}. 
Given either observational data or by inclusion of additional interventional data, clinical studies aim to either 1) quantify the causal effect of a treatment given the outcome \cite{hernan2020causal} or 2) infer the underlying causal structure of relationships between patient and treatment characteristics and relevant outcomes \cite{pearl2009causal,peters2017elements}. 

The former can be well formulated as the difference between the outcome expectations conditioned on different treatments (Average Causal Effect (ACE)) \cite{hernan2020causal}. A wide range of studies have built upon this methodology, including but not limited to target trial specification \cite{hernan2016specifying}, target trial emulation \cite{dickerman2019avoidable, caniglia2018emulating} and extending inferences from randomized trials to new target populations \cite{dahabreh2020extending}.

In comparison to the causal effect identification, we refer to 2) as (causal) structure identification \cite{peters2017elements}. The goal of structure identification is often to learn the entire causal structure, \ie a directed acyclic graph composed of nodes and edges that connect nodes. However, this is in general a non-deterministic polynomial-time (NP) hard problem \cite{zhu2019causal,luo2020causal}. To learn the cause-effect relations without inferring the entire causal structure, invariant causal prediction (ICP)~\cite{peters2016causal} was proposed to determine the set of causal variables given an outcome variable. Under the ICP framework, we utilize the concept `random variable' (informally, a function that assigns a set of possible samples to a measurable quantity) and define `causal variable' as follows (See Tab.~\ref{supp_tab1} in the supplement for summarized terminologies used in the paper): 

\begin{defi}
\label{ass:1}
Assume there exists a set of environments $u \in U$, 
given a collection of random variables $\bm{X}=(X_1, X_2, \ldots, X_n)$ and an outcome variable $Y$, if there exists a $\bm{X}_{S^*} = (X_{S^*_1}, \ldots, X_{S^*_j})$ with indices $S^* \subseteq \{1,\ldots,n\}$ such that
\begin{equation}
Y = f^*(\bm{X}_{S^*}^u) + \delta^u \; \forall u \in U, \; \mathrm{where} \; f^*:\mathbb{R}^{|S^*|}  \mapsto \mathbb{R}. 
\label{eq:inv}
\end{equation}
\begin{equation}
    \delta^u  \;\mathrm{are}
  \begin{cases}
    \text{identically distributed (i.d.)}       & \quad \text{if $\exists$ hidden confounders} \\
    \text{i.d. and $\delta^u \Perp \bm{X}_{S^*}^u$}  & \quad \text{else}
  \end{cases}
  \label{eq:confounder}
\end{equation}
\end{defi}
Then $\bm{X}_{S^*}$ are the \textbf{Plausible Causal Variables (under $U$)}. Here, $\bm{X}_{S^*}^u = (X_{S^*_1}^u, \ldots, X_{S^*_j}^u)$ are the corresponding random variables to $ \bm{X}_{S^*} = (X_{S^*_1}, \ldots, X_{S^*_j})$ created under the environment $u$.
For example, the (experimental) environment $u$ can arise via \textit{do intervention} \cite{pearl2009causal} ($do(X_0:=c)$) on $X_0$, then $X_0^u$ only samples the fixed value $c$. It is worth noting that the cause-effect relation $f^*$ in Eq.~\ref{eq:inv} stays invariant and independent of the environments $U$.
Next, we introduce the definition of \textbf{identifiable causal variables}.

\begin{defi}
\label{ass:2}
Following the specification of environments $U$, random variables $\bm{X}=(X_1, X_2, \ldots, X_n)$ and an outcome variable $Y$ in Definition~\ref{ass:1}, 
if there exists a $\bm{X}_{\bar{S}} = (X_{\bar{S}_1}, \ldots, X_{\bar{S}_j})$ with indices $\bar{S} \subseteq \{1,\ldots,n\}$ such that
\begin{equation}
\bar{S}:= \bigcap  \{ S \subseteq \{1,\ldots,n\} \: | \; \bm{X}_S \: \text{are plausible causal variables} \},
\label{eq:icv}
\end{equation}
then $\bm{X}_{\bar{S}}$ are the \textbf{Identifiable Causal Variables (under $U$)}, and are hence referred to as such. 
\end{defi}

Initially, vanilla ICP~\cite{peters2016causal} was presented and verified on linear cause-effect relations. Next, Heinze-Deml~\etal~\cite{heinze2018invariant} defined an environment $u$ as a random variable that is neither the descendant nor the parent of outcome variable Y, and conducted multiple conditional independence tests (CIT) on nonlinear cause-effect settings (NICP). Recently, Gamella~\etal~\cite{gamella2020active} suggested to investigate the stable set of variables instead, which is a relaxation of the set of identifiable causal variables (AICP). 
This progress in ICP has opened unprecedented paths to interpret complex datasets, especially the ones collected from RCTs. 
Finding which variables determine whether a treatment works or whether a patient will have a recurrence using ICP methods has great scientific potential. In application to the clinical domain, data interpretation by causal inference methods could improve our understanding of disease and aid in the design of new experiments and clinical trials. However, non-negligible efforts are required to adopt the existing ICPs to the clinical domain. This is due to 1) the complexity and multitude of variables that are considered relevant for treatment outcomes including patient level characteristics (patient demographics, text data from clinical records), information derived from images (radiology, pathology) and molecular data (genomic sequencing) and 2) the incompatibility between the error-tolerant implementation for the simulated dataset and safe-critical application relevant for medical decisions. Candidate ICP methods therefore need to be robust against noise and need to provide meaningful outputs that can be related to clinical risk in order to inform patient stratification. 

\begin{figure}[H]
\centering
\small
\includegraphics[width=1.\linewidth]{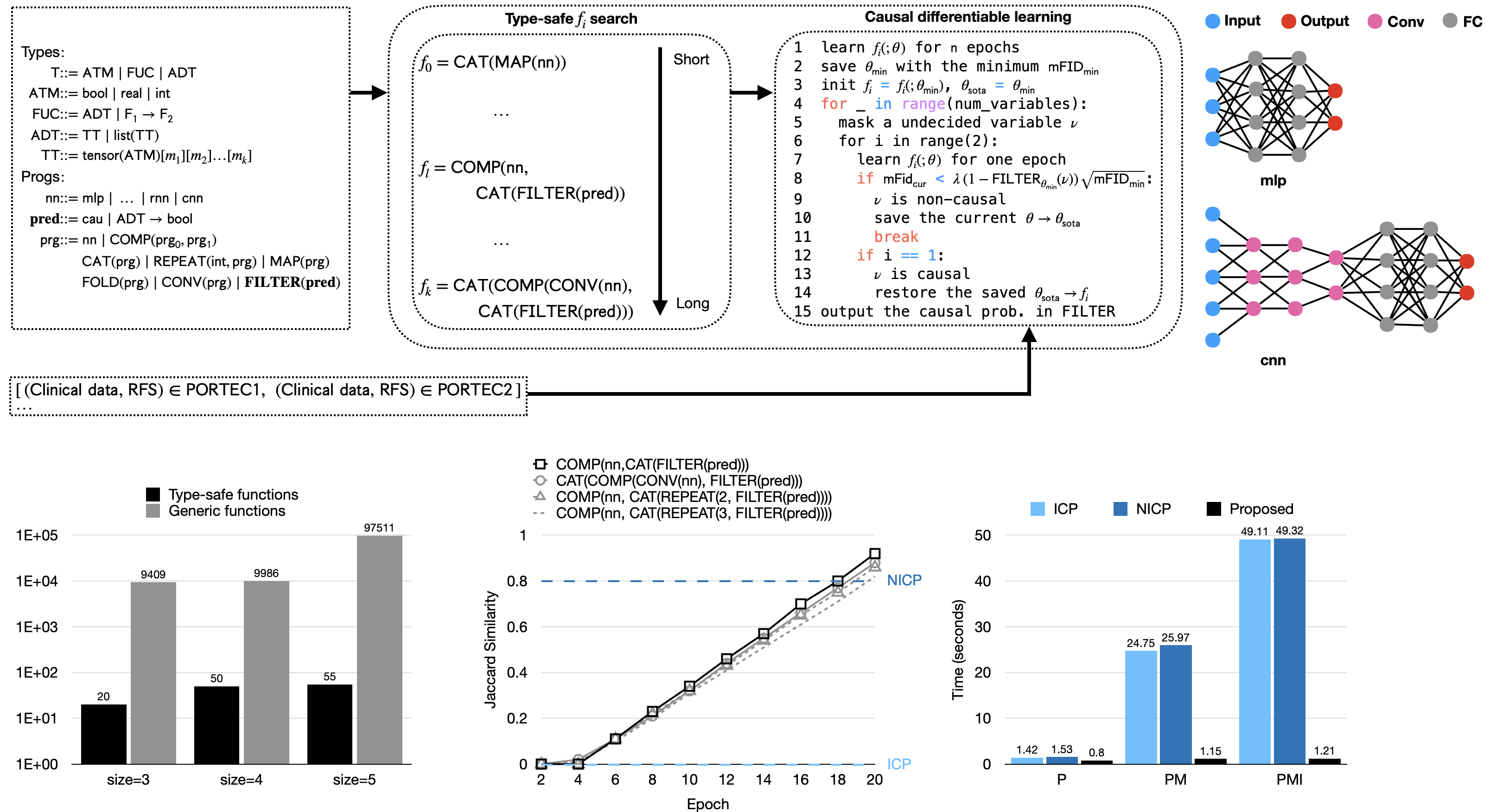}
\caption{The overall model illustration and performance of the proposed AutoCI.
Top: Illustrative scheme of the proposed AutoCI. Bottom left: The sampled numbers of type-safe functions vs generic functions. Here the size is the maximum amount of nn and pred functions allowed during the program synthesis. Bottom middle: The learning curve of the Jaccard Similarity (JS) for top-four type-safe functions achieved in the case with pathological, molecular and immune variables. 
Bottom right: The running time of determining the causal variables for pathological variables (P), pathological and molecular variables (PM), pathological, molecular and immune variables (PMI). Here the proposed AutoCI utilizes the function $\mathsf{COMP}(\mathsf{nn}, \mathsf{CAT}(\mathsf{FILTER}(\mathsf{pred})))$.}
\label{fig1}
\end{figure}

\section*{Results}
\label{sec:res}

\subsection*{Clinical variables overview}
The PORTEC 1 and 2 trials \cite{creutzberg2000surgery, nout2010vaginal} recruited 714 (since 1990 - 1997) and 427 (since 2000 - 2006) patients with early stage endometrial carcinoma respectively. 
305 cases from PORTEC 1 (42.7\%) and 335 cases from PORTEC 2 (78.5\%) with complete clinicopathological datasets were aligned and used in the experiments. 
Clinicopathologic characteristics of these subgroups were similar to the original trial populations (less than or equal to 17.3\% absolute difference in frequency of any variable). Importantly, there was no statistically significant difference in the variable of interest (mean and 5-year RFS) for causal variable identification between the excluded and included patient datasets, supporting that our analysis is representative of the overall study population (Supplementary Fig.~\ref{supp_fig1}, Tab.~\ref{supp_tab2}, \ref{supp_tab3}).
Considering PORTEC 1 and 2 as the two experimental environments, we aimed to determine the causal pathological, molecular and immune-related variables of EC recurrence status.

\noindent\textbf{Pathological variables (P).}
Pathological criteria tumor grading (Grade) \cite{scholten2005postoperative,stelloo2016improved}, lymphovascular space invasion (LVSI) \cite{bosse2015substantial} and myometrial invasion \cite{scholten2005postoperative,stelloo2016improved} are examined in the study, all of which are important indicators for an elevated risk of EC recurrence (See also the guideline \cite{concin2021esgo}). All variables were re-evaluated on formalin-fixed paraffin-embedded (FFPE) tumor material by specialized gynecopathologists to guarantee variable consistency for the two environments (trials). 

\noindent\textbf{Molecular variables (M).}
The molecular classification of endometrial cancer distinguishes four subtypes with validated prognostic impact: i) ultra-mutated EC with DNA-polymerase epsilon exonuclease domain mutations (POLEmut) with an excellent prognosis; ii) hypermutated EC with mismatch-repair deficiency (MMRd) with an intermediate prognosis; iii) copy-number-high EC with frequent TP53 mutations (p53abn) with an unfavorable prognosis; and iv) copy-number-low EC without a specific molecular profile (NSMP) with an intermediate prognosis \cite{kandoth2013mutational,stelloo2016improved}.
Pathogenic POLE mutations were detected by next generation sequencing of POLE hotspot exons \cite{church2015prognostic}. MMR and p53 status were determined by immunohistochemistry~\cite{stelloo2015refining}. Cases with more than one classifying feature were classified according to the dominant molecular feature on the basis of pathogenicity \cite{vermij2020incorporation}. Overexpression of L1CAM by tumor cells was assessed by immunohistochemistry using a cut-off of $>=10\%$ for positivity, and is associated with an increased risk of metastasis and death \cite{bosse2014l1,van2016prognostic}.

\noindent\textbf{Immune variable (I).}
(Intraepithelial) CD8+ T cell infiltration is an independent favorable prognostic indicator in early stage endometrial cancer \cite{horeweg2020prognostic}. To quantity CD8+ T cell infiltration in tissue samples of the PORTEC 1 and 2 trials, we compute CD8+ cell density derived from tissue microarrays (TMA) by immunohistochemistry and image analysis \cite{koelzer2019precision, horeweg2020prognostic}. Specifically, tissue microarrays capture cancer tissue samples from each patient in a highly standardized manner, allowing for the highly accurate evaluation of tumor and microenvironment-related factors in cancer samples for investigation with clinical outcomes \cite{zlobec2013next}.

Based on existing domain knowledge and biological understanding \cite{kandoth2013mutational,creutzberg2015nomograms,stelloo2015refining,karnezis2017evaluation,talhouk2019molecular,horeweg2020prognostic}, we consider the pathological ($\mathsf{P}$), molecular ($\mathsf{M}$) and immune ($\mathsf{I}$) variables to be the proxies of causal variables ($S_{\mathsf{prox}}$) (See Tab.~\ref{supp_tab3} in the supplement for more characteristics details).

\noindent\textbf{Sanity check variables.}
To investigate the robustness of the causal inference models, we intentionally include a randomized number as Patient ID 
and the vital tissue area of each TMA core (tissue area = sum of randomly sampled tumor and stroma areas from each case) in our subsequent analysis as non-causal variables. Based on prior domain knowledge, the variables "Patient ID" and "Tissue area" should not be causally related to cancer outcomes. The inclusion of these two non-causal variables thus serves as an important benchmark for comparison of the methods presented in this study. Importantly, as the two variables are expected not to impact clinical outcome and we attempt to verify that they do not impact the outcome, this strategy can also be considered as a typical use case of negative control \cite{lipsitch2010negative} in RCT studies.

\subsection*{PORTEC experiments overview}
As presented in Fig.~\ref{fig1} (top), our proposed AutoCI is composed of two key components: 1) A program synthesis language that searches the type-safe function candidates automatically; 2) A novel causal differentiable learning scheme that determines the causal variables. 
In the following, we first present the experimental results with emphasis on the individual components 1) and 2). We then demonstrate the overall performance on the main and ablation studies for the complete AutoCI method, confirming that the integration of program synthesis with a causal differentiable learning scheme is a critical step towards automated causal inference for clinical applications. 

\noindent\textbf{Automated type-safe function search.}
Type-safe functions are favorable candidates to strengthen the security and reliability of critical software applications \cite{yang2010safe}.  However, the manual verification of type-safe properties can be cumbersome, especially when examining thousands of function candidates.
Fig.~\ref{fig1} (bottom left) shows that the AutoCI can efficiently filter out a subset of promising type-safe differentiable functions (See also similar results in Valkov~\etal~\cite{valkov2018houdini}), and this can be done in a short period of time, $42.26s, 118.56s,  119.44s$ for $\text{size}=\{3, 4, 5\}$. By automatically excluding a large amount of generic functions that are not type-safe, it can greatly improve the development efficiency and algorithm safety compared to manual function design. After obtaining the type-safe candidates, we execute the causal differentiable learning scheme (Fig.~\ref{fig1} (top)) on the candidates and determine the set of predicted causal variables $S_{\mathsf{pred}}$. Here we utilize the Jaccard Similarity (JS)~\cite{gamella2020active} as the key metric to measure the prediction accuracy,
\begin{equation}
\label{eq:js}
\mathsf{JS}(S_{\mathsf{pred}}, S_{\mathsf{prox}}) = \frac{|S_{\mathsf{pred}} \cap S_{\mathsf{prox}} |}{ |S_{\mathsf{pred}} \cup S_{\mathsf{prox}}|}.
\end{equation}
Fig.~\ref{fig1} (bottom middle) demonstrates the JS accuracy with the growing epochs for the top-four type-safe candidates (See also Supplementary Tab.~\ref{supp_tab4}). We conclude that the function $f = \mathsf{COMP}(\mathsf{nn}, \mathsf{CAT}(\mathsf{FILTER}(\mathsf{pred})))$ achieves the optimal JS score, \eg $91.9 \pm 0.06\%$, and thus is used as the default function for further analysis. As pointed out in Valkov~\etal~\cite{valkov2018houdini}, HOUDINI allows us to transfer high-level modules across learning tasks. More specifically,
the type-safe candidates discovered via the search algorithm are agnostic of disease-specific features for hazard analysis.
Independent of the hazard analysis conducted for cancer studies, 
these type-safe candidates can therefore be re-used and fine-tuned to perform causal variable identification given data on the survival outcome for each patient. 
Depending on the JS score achieved by the candidates we determine the optimal learned type-safe model.
As a result, the proposed AutoCI approach can pave the way towards an efficient causal analysis for many of the real-world cancer studies.

\begin{table}
\centering
\includegraphics[width=1.\textwidth]{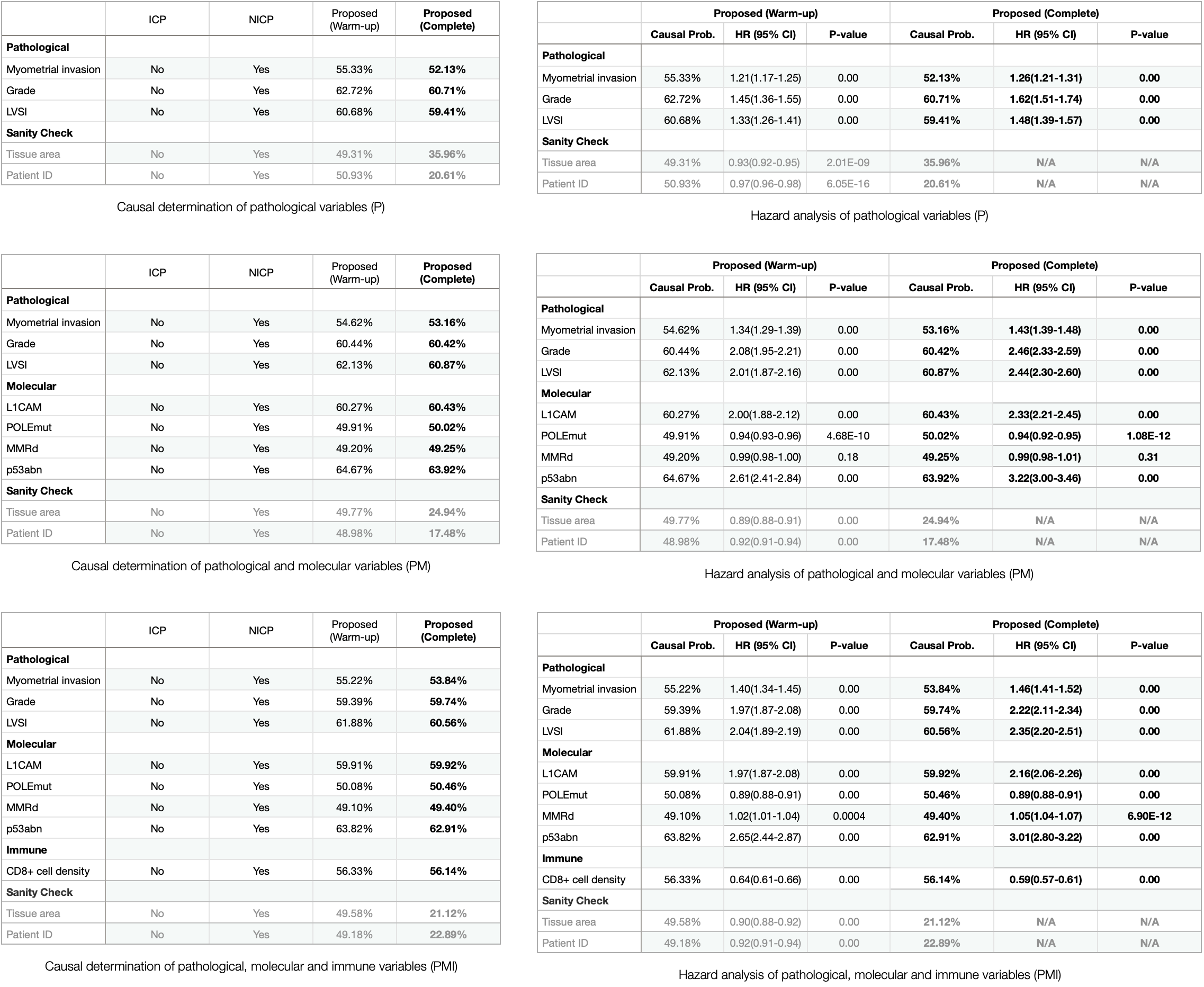} 
\caption{The comparison of causal variable determination for the PORTEC dataset among ICP, NICP and the proposed method.
Left column: The causal variable determination of P, PM and PMI. Here, we report Yes (Causal) or No (Non-causal) for ICPs and causal probability for the proposed AutoCI. Right column: The hazard analysis including the results of Hazard ratio (HR), 95\% Confidence interval (CI) and p-value for P, PM and PMI, where the p-value is computed from the chi-square test. The Proposed (Warm-up) refers to the Step 1-2 of the pseudo code in Fig.~\ref{fig1}.} 
\label{tab1} 
\end{table} 

\noindent\textbf{Determining causal variables with clear differentiation.}
We compare the AutoCI with the state-of-the-art ICP methods, ICP and NICP. Although competitive results are achieved by AICP on the toy experiments, AICP requires the re-generation of additional interventional data in each learning step. Such a learning scheme is incompatible to the real-world RCTs setting, hence AICP is not applicable (N/A) for the PORTEC experiments. 
As ICP and NICP explicitly accept or reject the variable of interest, we report YES or NO in Tab.~\ref{tab1}. For the proposed AutoCI, we report the mean of the causal probabilities for each variable.
Overall, the proposed AutoCI outperforms the ICP and NICP in terms of differentiating causal and non-causal variables ($\geq 50\%$ vs $< 25\%$ for PMI), demonstrating its advantages over the SOTA methods with a clear margin (See Tab.~\ref{tab1} (left column)). When examining the individual variables of interest, we can see that ICP fails to determine the proxy variables to be the causal ones, while for NICP all the variables including the sanity check ones are considered to be causal. 
These results clearly contrast to the methodological comparison of the ICP methods on the toy data respecting the normal distributions. If we decompose the proposed causal learning scheme, we witness the suppression of the causal probability on the sanity variables over the warm-up stage (Step 1-2 of the pseudo code in Fig.~\ref{fig1}), while the causal variables do not show deterioration of performance. This clear differentiation of causal and non-causal variables can aid the definition of meaningful cut-offs by AutoCI on a given cohort guided by clinical expertise.

\noindent\textbf{Hazard analysis of the individual variables.} In parallel to the causal variable determination, the corresponding hazard ratio (HR) analysis on endometrial cancer (EC) recurrence is also performed.
In the scenario when unknown spurious (non-causal) variables are included in the hazard analysis, the causal cut-off can help reducing the noise introduced by non-causal variables. For instance, Tab.~\ref{tab1} (right column) reports the decreased hazard of Tissue area (0.90 (0.88-0.92)), Patient ID (0.92 (0.91-0.94)) achieved in the warm-up stage for PMI case. Without causal analysis, one may falsely conclude that larger tissue area leads to a slightly lower risk of cancer recurrence.  
Besides, the hazard ratio achieved within the warm-up and complete learning stage remains stable and consistent to the standard clinical interpretation, \ie the values assigned to the variable indeed correspond to either poor or favorable outcome correctly. Therefore, this learning scheme can help delivering reliable outputs that are understandable for clinical experts.

\begin{figure}[H]
\centering
\small
\includegraphics[width=1.\linewidth]{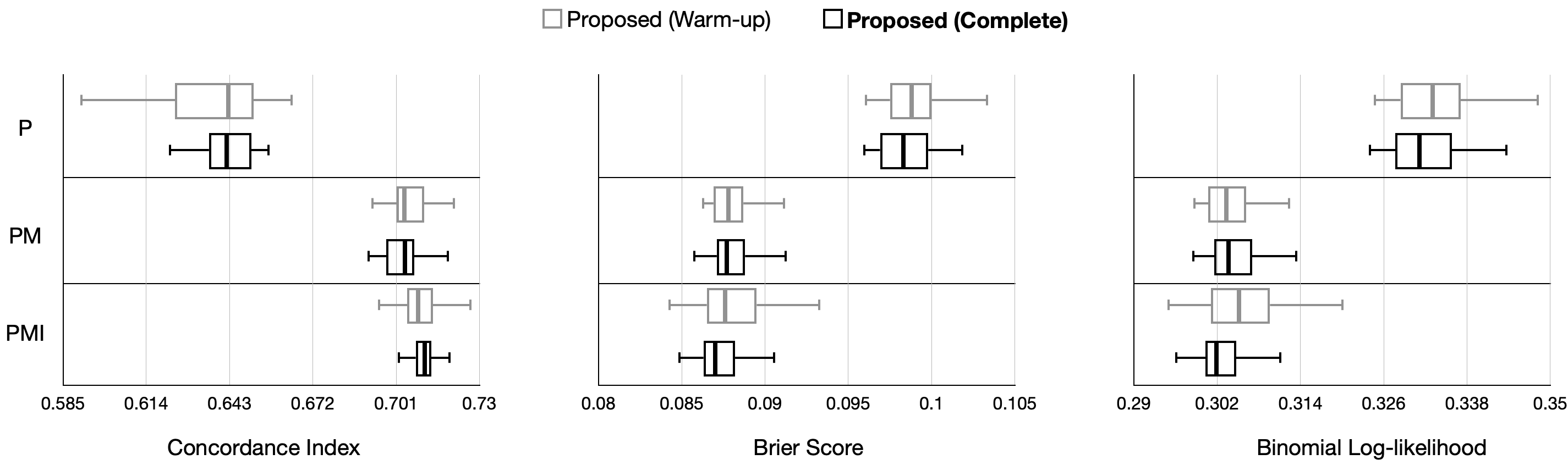}
\caption{The evaluation metrics of hazard analysis conducted on the PORTEC dataset. 
The box plots of Concordance Index (Left), Brier Score (Middle), and Binomial Log-likelihood (Right) that are derived from n=640 patients included in the PORTEC dataset, where the box bounds the interquartile range (IQR) divided by the median, and whiskers extend to $\pm 1.5\times$IQR beyond the box.}
\label{fig2}
\end{figure}

\noindent\textbf{Ablation study with hidden confounders.}
To elaborate the robustness of AutoCI, we conducted ablation studies with the influence of confounding. This is achieved by step-wise inclusion of pathological variables (P), pathological and molecular variables (PM). As shown in Tab.~\ref{tab1} (top and middle rows), the proposed AutoCI presents consistent advantages over the existing ICP methods in terms of learning meaningful causal probabilities for both non-causal and causal variables. For instance, the Tissue area, Patient ID are determined to be 35.96\%, 20.61\% for P and 24.94\%, 17.48\% for PM, while the causal probability of all the proxy variables remain close to or above 50\%. In the hazard analysis, the results of the confounding studies P and PM are also consistent with the results reported for the main study (PMI). For instance, L1CAM and p53abn are assigned with increased hazards for PM and PMI, indicating poor prognostic outcome. These results are in agreement with clinical understanding \cite{kandoth2013mutational, bosse2014l1, stelloo2016improved}. 
The numerical improvements from P to PMI in the accuracy of probabilistic predictions (See Fig.~\ref{fig2}) provide complementary evidence confirming the robustness of AutoCI. When comparing to the warm-up stage of AutoCI, we further observe a reduction in variance in the evaluation of the complete causal-aware AutoCI. 
Lastly, the running time of the P, PM and PMI studies grows only mildly for the proposed method, in contrast to the dramatic increase in time complexity for ICP and NICP, where the bottleneck of the ICPs lies in the exponential increase $o(2^{|S|})$ in search space with the growing number of variables $S$ (Eq.~\ref{eq:icv}).  

\section*{Discussion}
In this study, we proposed a novel automated causal inference algorithm (AutoCI). Taking two large RCTs as the experimental environments, we re-interpret the clinical variables of interest from a causal perspective.
Compared to the existing ICP methods, the proposed AutoCI demonstrates consistent advantages in determining causal variables with the presence of hidden confounders. Complementary to the standard hazard analysis, AutoCI provides an automatic tool for medical data analysis to investigate causal association of patient variables from a new perspective, offering informative and critical evidence to support clinical interpretation. Specifically, the accurate determination and exclusion of the spurious (non-causal) variables is a key step to enable more precise patient stratification in the future. 

\noindent\textbf{Design choices of AutoCI.}
Dissimilar to generic algorithms, clinical algorithms must deal with unique challenges in terms of ensuring the safety \cite{allen2019role} and robustness. Error-prone algorithms can potentially lead to critical errors in medical care. Driven by the need to develop safe-critical applications, AutoCI is carried out with a type-safe program synthesis method~\cite{valkov2018houdini}. By further incorporating a newly proposed causal-aware module into this framework, we are able to synthesize a subset of differentiable type-safe candidates well suited for causal-aware learning. Compared to the laborious and error-prone manual function design, this implementation improves the efficiency and safety of AutoCI. Moreover, to achieve the robustness on the real clinical tasks, we introduce a novel causal differentiable learning scheme that utilizes the Fr\'echet inception distance (FID) \cite{heusel2017gans}. As a whole, the proposed AutoCI is the seamless integration of both components. 

\noindent\textbf{Comparison to existing ICPs.}
Application of the prior ICP methods has confirmed the feasibility of causal variable learning on toy experiments (See Tab.~\ref{tab2}). This is substantiated by the outstanding results in the absence of confounders. However, the error-tolerant implementations of prior ICPs on the synthesized experiments are not well-tailored for real clinical applications, especially in the presence of hidden confounders. Compared to the ICP, AICP and NICP, the AutoCI presents robust results on both toy and PORTEC experiments. With the inclusion of confounders, AutoCI demonstrated a robust differentiation between causal and non-causal variables for PORTEC, and achieves superior quantitative scores on both the finite sample and ABCD settings. 

\noindent\textbf{Clinical interpretations.}
Importantly, the hazard analysis and ranking of clinicopathological and molecular variables using AutoCI (See Tab.~\ref{tab1}) is generally consistent with the common biological and clinical interpretation. Taking pathological variables as an example, studies~\cite{scholten2005postoperative,stelloo2016improved,bosse2015substantial} indeed show that grade, deep myometrial invasion and LVSI are significant independent predictors of early EC recurrence. The causal probabilities provided by AutoCI thereby give additional information on the relevance of each variable for the determination of outcome, and the likelihood of each variable is consistent with domain expertise. LVSI is considered to be a significant predictor independent of molecular subgroup and is ranked with the highest causal probability, while grade and myometrial invasion are indeed weaker but independent prognostic indicators.

Further, AutoCI correctly identifies the prognostic associations of the molecular variables of endometrial cancer, assigns the appropriate hazards for outcome, and ranks the molecular subgroups in the order of causal probability that would be expected by expert's domain knowledge. Specifically, the molecular factors with the highest adverse risk, p53 abnormality (3.01 (PMI), 3.22 (PM)) and L1CAM over-expression (2.16 (PMI), 2.33 (PM)) \cite{bosse2014l1} are recognised as such, while the POLEmut is consistently associated with a reduced risk of disease relapse as confirmed in previous studies \cite{church2015prognostic}. Adding an immune variable further refines the model, as expected by domain expertise \cite{horeweg2020prognostic}, and highlights a causal relationship between cytotoxic T-cell responses and EC recurrence in early-stage endometrial cancer. In summary, AutoCI correctly quantifies and ranks causal pathological, molecular and immune variables for patient outcomes in the clinical trial setting.

Going beyond academic toy models, the proposed AutoCI extends the researchers current statistical toolbox with a new causal-driven method that can assign the causal likelihood to prognostic and predictive variables. As such, this method will enable identification of clinically relevant variables among the ever-increasing number of biomarkers in cancer research that show statistical correlation with clinical outcome. Hence, while the direct real-world application of this method is primarily scientific, subsequent clinical validation and development may enable better selection of (bio)markers to stratify patients for cancer treatments and prediction of prognosis.

\noindent\textbf{Limitation.}
Despite the clear cut-off between non-causal and proxy variables provided by AutoCI, some of the proxy variables present borderline hazard ratios, for instance MMRd. Due to small effect size, clinical studies \cite{smyth2017mismatch,leon2020molecular} usually associate  Mismatch Repair Deficiency (MMRd) with intermediate patient prognosis, not significantly different from the prognosis of NSMP EC. However, from the biological perspective, MMRd is highly relevant for a well-defined cascade of molecular changes in cancer cells with favorable prognostic impact as proven by a large number of well-designed experimental and translational studies \cite{stelloo2015refining,kloor2018immune}. The loss of DNA mismatch repair capability in cancer cells leads to a strong increase in tumor mutational burden caused by mismatch, frameshift and insertion/deletion mutations. Due to the structure of eucaryotic DNA, frameshift mutations frequently lead to the translation of truncated peptides that are highly immunogenic, and contribute to the induction of an effective anti-tumoral immune response \cite{kloor2018immune}. In consistency with biological understanding, AutoCI identifies MMRd as causally related to outcome in the present study, though the lack of a statistically significant association with EC recurrence requires further investigation.

\noindent\textbf{Conclusion.}  In this study, we investigate the causal variable determination among multiple RCTs and present its advantages over the compared methods on both toy and PORTEC experiments. For clinical application, further validation of this methodology in independent clinical trial datasets will be needed to ensure generalisation.

\begin{table}[H]
\centering
\small
\includegraphics[width=0.5\textwidth]{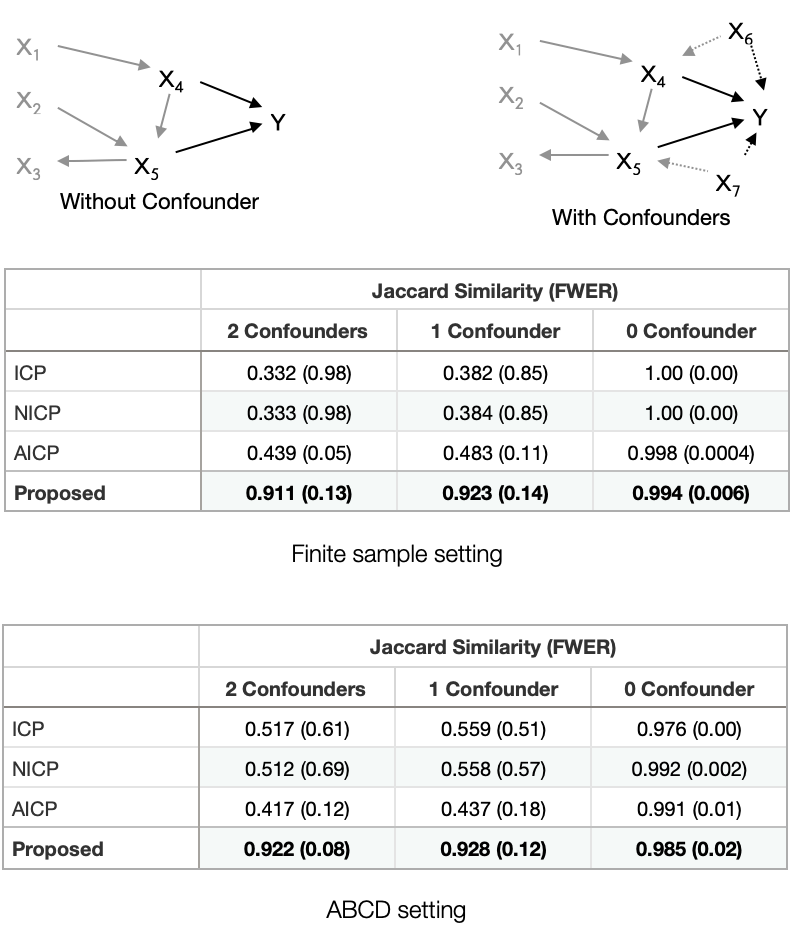} 
\caption{The comparison of causal variable determination for the toy datasets between ICPs and the proposed method.
Top: The illustration of the Structural Causal Model (SCM), from which the data are sampled for the cases of finite sample and ABCD setting. Middle: The results of the compared methods for the Finite sample setting. Bottom: The results of the compared methods for the ABCD setting. } 
\label{tab2} 
\end{table} 

\section*{Methods}
\label{sec:met}
As embedded in the AutoCI abbreviation, automated and causal are the two building blocks of the proposed method. 
Concretely, the automation component is implemented with a type-safe program synthesis language HOUDINI \cite{valkov2018houdini}. Further, we introduce a novel differentiable causal learning scheme that is built upon ICP.

\subsection*{Type-safe program synthesis}
HOUDINI \cite{valkov2018houdini} is a typed language with a rich set of pythonic higher-order functions such as MAP, FOLD, COMP (Pythonic: $\mathsf{map()}, \mathsf{reduce()}, \mathsf{lambda \; x: f(g(x))}$), etc (See Fig.~\ref{fig1} (top left)).
Relying on the built-in method for program search, it allows us to efficiently search promising type-safe differentiable program candidates. Compared to other program synthesis languages \cite{gaunt2017differentiable,mao2018neuro,vedantam2019probabilistic,ellis2020dreamcoder}, HOUDINI rules out the error-prone functions that undermine the software safety and presents itself as an ideal candidate for our task (See Supplementary Tab.~\ref{supp_tab5} for further comparison). 

Despite of rich built-in functions provided by the HOUDINI, it lacks an explicit flow control mechanism. Driven by the need of integrating the causal-aware learning, we introduce the predicate module ($\bm{\mathsf{PRED}}$) containing the function (cau) (See also Fig.~\ref{fig1} (top left)),

\begin{equation}
\label{eq:cau}
\mathsf{cau}(\bm{x}; \bm{\theta}) = \mathsf{mask} \odot \mathsf{sigmoid}(\bm{\theta}) \odot \bm{x},
\end{equation}
where $\bm{\theta}$ are the learnable weights (normalized by $\mathsf{sigmoid}$), $\odot$ is the element-wise multiplication, $\mathsf{mask}$ is the vector containing $0$ or $1$ manipulated in Step 5 of Fig.~\ref{fig1} (top), $\mathsf{mask} \odot \mathsf{sigmoid}(\bm{\theta})$ presents the causal probability for each variable. Together with the newly introduced higher-order function $\bm{\mathsf{FILTER}}$, we are able to synthesize type-safe causal-aware programs.

\subsection*{Causal differentiable learning}
In Definition~\ref{ass:1}, the ICP does not make assumptions on the function $f^*$ (Eq.~\ref{eq:inv}). In real-world applications, it is reasonable to specify the search space of $f^*$. If we assume that $f^*$ is differentiable, then we have a trivial extension:
\begin{equation}
\label{eq:cau1}
\begin{aligned}
f:& \quad \mathbb{R}^{n} \mapsto \mathbb{R} \\
  & \quad \underbrace{X_{i_0}, \ldots, X_{i_{|S^*|}}}_{\bm{X}_{S^*}}, 
    \underbrace{X_{i_{|S^*| + 1}}, \ldots, X_{i_n}}_{\bm{X}_{S^{*c}}}
   \rightarrow f^*(\bm{X}_{S^*}),  \quad \forall u \in U,
\end{aligned}
\end{equation}
where $f$ remains differentiable \wrt all the variables $\bm{X}$. More importantly, the gradient norms \wrt the non (plausible) causal variables $\bm{X}_{S^{*c}}$ should vanish, \eg $\|\nabla_{S^{*c}}f\|=0$.
Motivated by the extension, we first assume $f^*$ to be differentiable in Definition~\ref{ass:1}, then we have the claim:
\begin{clm}
\label{clm:1}
Following the specification of environments $U$, random variables $\bm{X}=(X_1, X_2, \ldots, X_n)$ and an outcome variable $Y$ in Definition~\ref{ass:1}, if $\bm{X}_{\hat{S}} = (X_{\hat{S}_1}, \ldots, X_{\hat{S}_j})$ with indices $\hat{S} \subseteq \{1,\ldots,n\}$ are the identifiable causal variables, then there exists a differentiable function $f(\bm{X}):\mathbb{R}^{n} \mapsto \mathbb{R}$ satisfying (\ref{eq:inv}, \ref{eq:confounder}) such that $f$ has the maximum amount $|\hat{S}^c|$ of variables with $\|\nabla_{\hat{S}^c}f\|=0$.
\end{clm}
From this perspective, we can reduce the ICPs to learning an invariant differentiable function $f$, where $f$ has the (maximum amount of) vanishing gradient norms on the non-causal variables. Such reduction enables us to smoothly integrate the ICP into the modern differentiable learning framework. 

\noindent\textbf{Algorithm design.} To impose the vanishing gradient norms, we seek for the $\mathsf{mask}$ vector in Eq.~\ref{eq:cau} as the solution.
Initially, we assign $\mathsf{mask} = 1$ for all the variables. Assume we mask a causal variable $X_i$ by flipping $\mathsf{mask_i} = 0$, then it should greatly disturb the learning errors across multiple environments. If this is the case,  we restore $\mathsf{mask_i} = 1$ and take $X_i$ as the causal variable. Otherwise we reject the variable $X_i$ and $\mathsf{mask_i}$ remains to be $0$, which imposes the zero gradient with respect to $X_i$. 
To quantify the disturbance when variables of interest are missing, existing ICPs use several statistical tests \cite{pfanzagl1996studies,levene1961robust,wilcoxon1992individual}. These tests suffer from capturing the nuance of distributions related to higher-dimensional clinical data. Motivated by the recent success in complex vision data \cite{lucic2017gans}, we utilize the Fr\'echet inception distance (FID) \cite{heusel2017gans} that is derived from the Wasserstein distance \cite{villani2008optimal}. More specifically, the square root of FID between Gaussian distributions is exactly $W_2$ Wasserstein distance \cite{villani2008optimal}, \ie satisfying three axioms: identity, symmetry, triangular inequality, 
whereas the statistical tests used in ~\cite{peters2016causal,heinze2018invariant,gamella2020active} are generally not mathematical metrics.
As a result, the maximum FID (mFID) of different environments $U$ is proposed to measure the distribution difference,
\begin{equation} 
\mathsf{mFID} = \max_{u \in U} \mathsf{FID}\,(\bm{\mu}_u, \bm{\mu}_{u^c}),
\end{equation}
where $\bm{\mu}_u, \bm{\mu}_{u^c}$ are the distributions w.r.t the data sampled from the environment(s) $\{u\}$ and $\{u\}^c$.
In the supplement, Tab.~\ref{supp_tab6} presents side by side comparisons between $\mathsf{mFID}$, F-test + t-test ~\cite{peters2016causal,gamella2020active} and  Levene-test + Wilcoxon-test ~\cite{heinze2018invariant}, all of which are applied for training the same type-safe function $\mathsf{COMP}(\mathsf{nn}, \mathsf{CAT}(\mathsf{FILTER}(\mathsf{pred})))$ under the proposed causal differentiable learning scheme. As displayed in the supplementary Tab.~\ref{supp_tab6}, we conclude that the proposed $\mathsf{mFID}$ outperforms the compared statistical tests with a clear margin.
For the pseudo code of proposed algorithm please check the top plot of Fig.~\ref{fig1} (See the box of `causal differential learning' for more details).

\begin{figure}[H]
\centering
\small
\includegraphics[width=1.\linewidth]{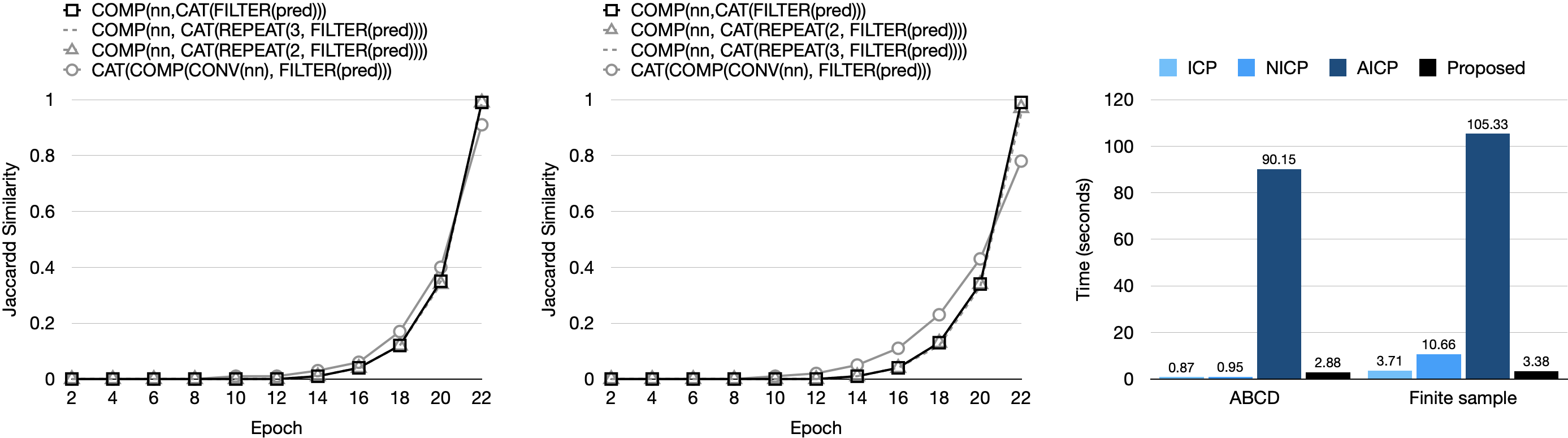}
\caption{The learning performance and time of the optimal type-safe $f = \mathsf{COMP}(\mathsf{nn}, \mathsf{CAT}(\mathsf{FILTER}(\mathsf{pred})))$ on toy datasets. Left: The learning curve of the Jaccard Similarity(JS) for top-four type-safe functions (Finite sample setting). Middle: The learning curve of the Jaccard Similarity(JS) for top-four type-safe functions (ABCD setting). Right: The running time of the compared methods for the Finite sample and ABCD settings. Here the proposed AutoCI utilizes the function $\mathsf{COMP}(\mathsf{nn}, \mathsf{CAT}(\mathsf{FILTER}(\mathsf{pred})))$.}
\label{fig3}
\end{figure}

\subsection*{Proof of concept}
For the sake of concept validation, we first conduct experiments on toy datasets. We compare the proposed AutoCI to the SOTA methods ICP, NICP and AICP. 
Specifically, we follow the two experimental protocols presented in AICP \cite{gamella2020active}: Finite sample setting and ABCD setting \cite{agrawal2019abcd}. The former presents the ideal scenario where the same amount of data (1000) are sampled from both observational and experimental (interventional) environments, the latter simulates a more realistic case where limited experimental data (10) are collected in conjunction with a large amount of observational data (1000). 
The data of both settings are generated from randomly chosen linear structural causal models (SCM). In our experiments, 400 SCM models are tested to guarantee the reliability of our results.
For the compared ICP methods, we applied the optimal strategies discussed in the paper and parameters are fine-tuned to the experiments. Specifically, careful parallelization and code optimization is also performed for ICP methods. We use 16 cores of CPU Intel(R) Core(TM) i7-7820X CPU @ 3.60GHz to train the ICP methods in parallel and the GPU NVIDIA TITAN V (12GB) to train the AutoCI.   
For the proposed AutoCI, we use the standard Adam optimization \cite{kingma2014adam} with the learning rate $0.02$ throughout the experiments. For the warm-up stage (Step 1-2 of Fig.~\ref{fig1}) of causal differentiable learning we adopt eight epochs. 
The batch size is set to be $64$ for all the experiments.
We calibrate the $\lambda = 5, 1$ for toy and PORTEC on a small subset of unused data. For the toy experiment, we apply the MSE loss to supervise the learning process and report the result obtained by training the AutoCI one time, where the non-causal variable is determined to be the one with $0$ causal probability (Eq.~\ref{eq:cau}). 
For the PORTEC experiment, we utilize the partial likelihood to learn the hazard coefficient \cite{kvamme2019time}. To fully utilized the PORTEC patient data and incorporate into the differentiable cox model \cite{kvamme2019time}, the molecular subtype variables POLEmut, MMRd and p53abn are assigned with 1 if present else (including NSMP) 0.  
To guarantee the representativeness of the PORTEC results, we independently train the AutoCI 64 times and average the causal probability for each variable. Complementary to JS score, we also report the Family-wise error rate $\mathsf{FWER} = P(S_{\mathsf{pred}} \nsubseteq S_{\mathsf{prox}})$ (type-I error).   

As shown in Tab.~\ref{tab2}, our AutoCI achieved competitive JS and FWER scores compared to the ICP methods. Clearly, the proposed method is more resistant to the influence of hidden confounders and all the results reach $> 90\%$ JS accuracy. This is achieved by the optimal type-safe function $\mathsf{COMP}(\mathsf{nn}, \mathsf{CAT}(\mathsf{FILTER}(\mathsf{pred})))$ (Fig.~\ref{fig3} (left and middle), Supplementary Tab.~\ref{supp_tab7}, \ref{supp_tab8}).
Such advantages also confirm the effectiveness of the proposed causal learning scheme with the utilization of mFID metric. 
Similar to the PORTEC experiments, because of the exhaustive subset research required in Eq.~\ref{eq:icv}, the time complexity of ICP and NICP raises dramatically from ABCD to Finite sample settings (Fig.~\ref{fig3} (right)).

\section*{Data availability}
The code used to generate the data of the toy experiments is available via the link \url{https://github.com/juangamella/aicp}.
TransPORTEC complies with the FAIR guiding principles for scientific data management and stewardship. TransPORTEC is oriented towards a knowledge and data sharing policy. In other words, expertise and structure of different groups will be made available for all the proposals and the projects presented by the member groups. In the knowledge-sharing policy of the consortium, the possibility and the power offered by the collected databank of samples and data can be used by every participating member. All member researchers/groups are welcome to introduce new proposals. In principle, all the anonymized pathological and molecular data already included in the TransPORTEC biobank will be available for every member of the consortium upon request. In the future, after completion of the primary joint projects, the biobank data will be opened towards use by researchers worldwide upon request.

\section*{Code availability}
Our code is implemented with PyTorch and publicly accessible via \url{https://github.com/CTPLab/AutoCI}, which is released under the MIT licence.

\section*{Acknowledgments}
The authors would like to convey the gratitude to all clinicians and technicians participated in the PORTEC 1 and 2 trials (registration number ISRCTN16228756), and all scientists, pathologists and patients involved in the data processing and analysis. 
The PORTEC 1 and PORTEC 2 trials were supported by the grants from the Dutch Cancer Society (CKTO 90–01 and CKTO 2001–04, respectively).
The molecular profiling was supported by the grants from the Dutch Cancer Society (KWF UL2012-5447 and KWF/YIG 10418, respectively). 
Prof. Koelzer reports grants from Promedica Foundation (F-87701-41-01) during the conduct of the study.
Dr. Horeweg reports grants from the Dutch Cancer Society (KWF-2021-13400, KWF-2021-13404) during the conduct of the study.

\section*{Author Contributions}
J.W. and V.H.K. conceived the research idea. 
J.W. implemented the algorithm and ran the experiments. 
J.W., V.H.K., and N.H. analysed the data and results. 
J.W. and V.H.K. wrote the manuscript. 
N.H. and V.H.K. reviewed the manuscript and supervised this study.
R.A.N., I.M.J.-S., J.J.J., L.C.H.W.L., E.M.v.d.S.-B., T.B., C.L.C., M.d.B., H.W.N., T.B., and V.T.H.B.M.S. provided the data (PORTEC1 and PORTEC2 trials).
C.L.C. and V.T.H.B.M.S. conceptualized and designed the PORTEC 1 and PORTEC 2 study.
R.A.N., T.B., C.L.C., and V.T.H.B.M.S. supervised the PORTEC 1 and PORTEC 2 study.

\section*{Competing interests}
The authors declare no competing interests.

\section*{Ethical approval}
The PORTEC study protocols were approved by the Dutch Cancer Society and by the medical ethics committees at participating centers. Both studies were conducted in accordance with the principles of the Declaration of Helsinki. All patients provided informed consent for study participation. The PORTEC 1 trial was registered at the Daniel Den Hoed Cancer Center (DDHCC) Trial Office. The PORTEC 2 trial was registered in ClinicalTrials.gov under the identifier NCT00376844.

\newpage
\noindent\textbf{\large Supplementary material}

\begin{table}[h]
\center
\begin{tabular}{ll}
\hline
\textbf{Abbreviation} & \textbf{Definition} \\  \toprule
RCT                   & Randomised controlled trials \\  
PORTEC                & Post operative radiation therapy in endometrial carcinoma \\  
EC                    & Endometrial carcinoma (cancer)   \\
ESGO                  & European Society of Gynaecological Oncology    \\
ESTRO                 & European Society for Radiotherapy and Oncology    \\
ESMO                  & European Society of Medical Oncology    \\ \hline
Grade                 & Tumor grading \\
LVSI                  & Lymphovascular space invasion \\
POLEmut               & Polymerase epsilon mutant EC \\
MMRd                  & Mismatch repair deficient EC \\
p53abn                & p53 abnormal EC \\
NSMP                  & EC with no specific molecular profile \\
L1CAM                 & L1 cell adhesion molecule \\
P                     & Pathological variables \\
PM                    & Pathological and molecular variables \\
PMI                   & Pathological, molecular and immune variables \\
HR                    & Hazard ratio \\ 
CI                    & Confidence interval \\
i.d.                  & identically distributed \\\toprule
NP                    & Non-deterministic polynomial-time \\
SCM                   & Structural causal model \\
ICP                   & Invariant causal prediction \\
PRED                  & Predicate module \\
FID                   & Fr\'echet inception distance \\
JS                    & Jaccard similarity \\
FWER                  & Family-wise error rate \\
ABCD                  & Active budgeted causal design strategy
\end{tabular}
\caption{The abbreviation table of clinical, statistical and causal definitions.}
\label{supp_tab1}
\end{table}

\begin{figure}
\centering
\includegraphics[width=0.6\textwidth]{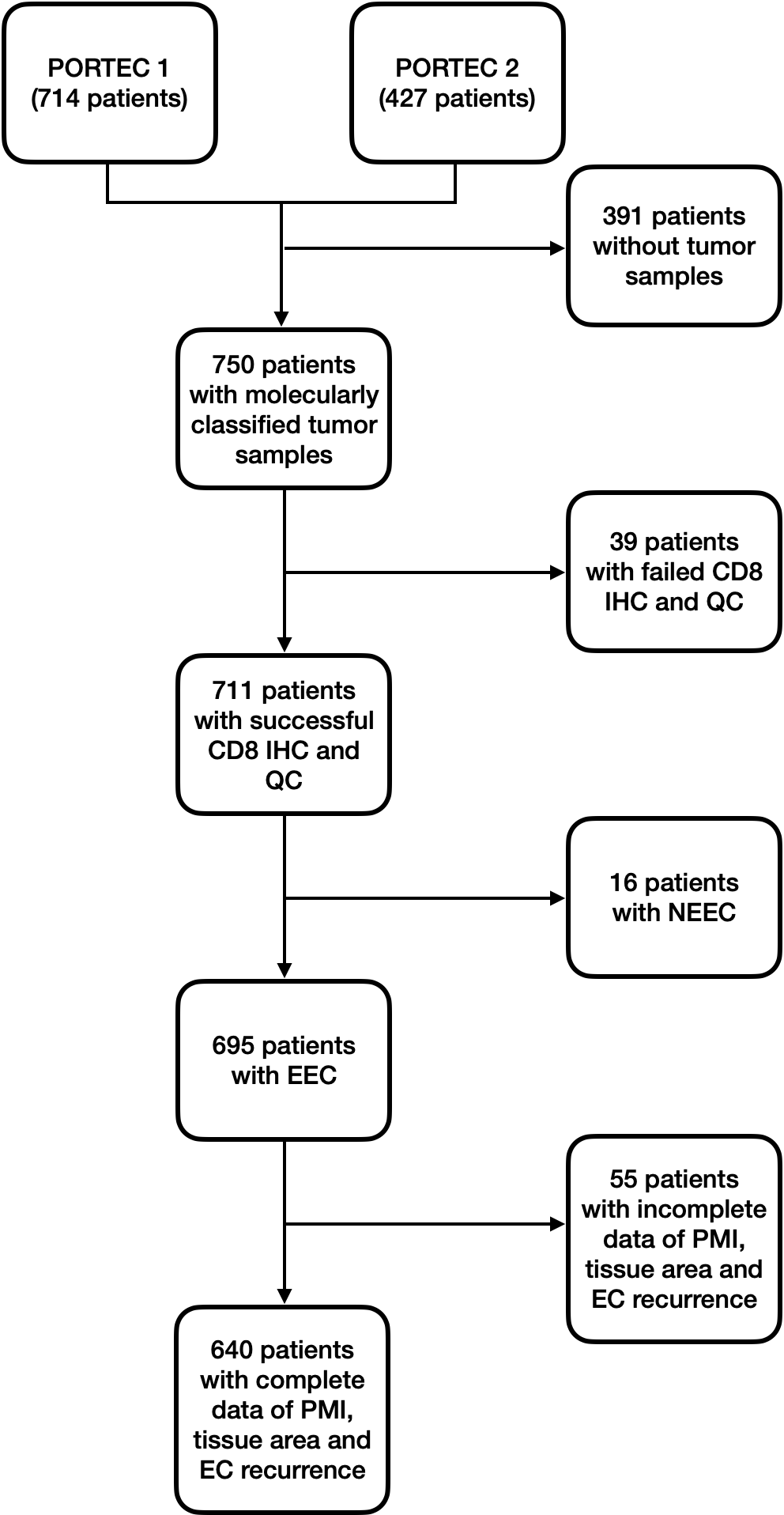} 
\caption{The consort diagram presenting the process of patient selection. Abbreviations: QC -  quality control, IHC - immunohistochemistry, EEC- endometrioid endometrial carcinoma, NEEC- non-endometrioid endometrial carcinoma.} 
\label{supp_fig1} 
\end{figure} 

\begin{table}
\centering
\includegraphics[width=0.85\textwidth]{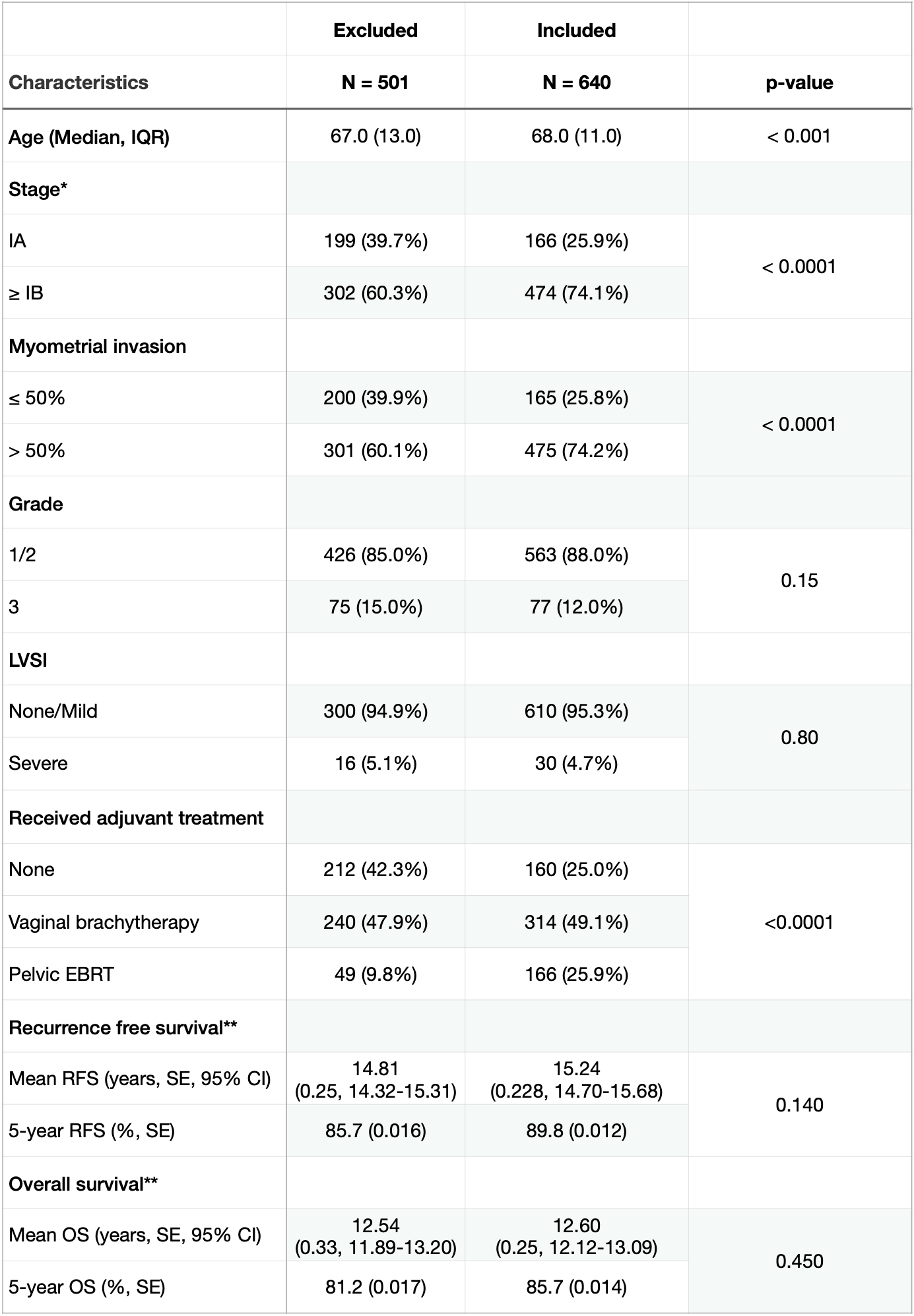} 
\caption{The characteristics comparison of excluded and included patients. * After a posteriori central review 2 cases were classified as stage II, 2 as stage IIIA and 1 as stage IIIB. ** The p-values of RFS and OS are computed from log-rank test.} 
\label{supp_tab2} 
\end{table}

\begin{table}
\centering
\includegraphics[width=0.7\textwidth]{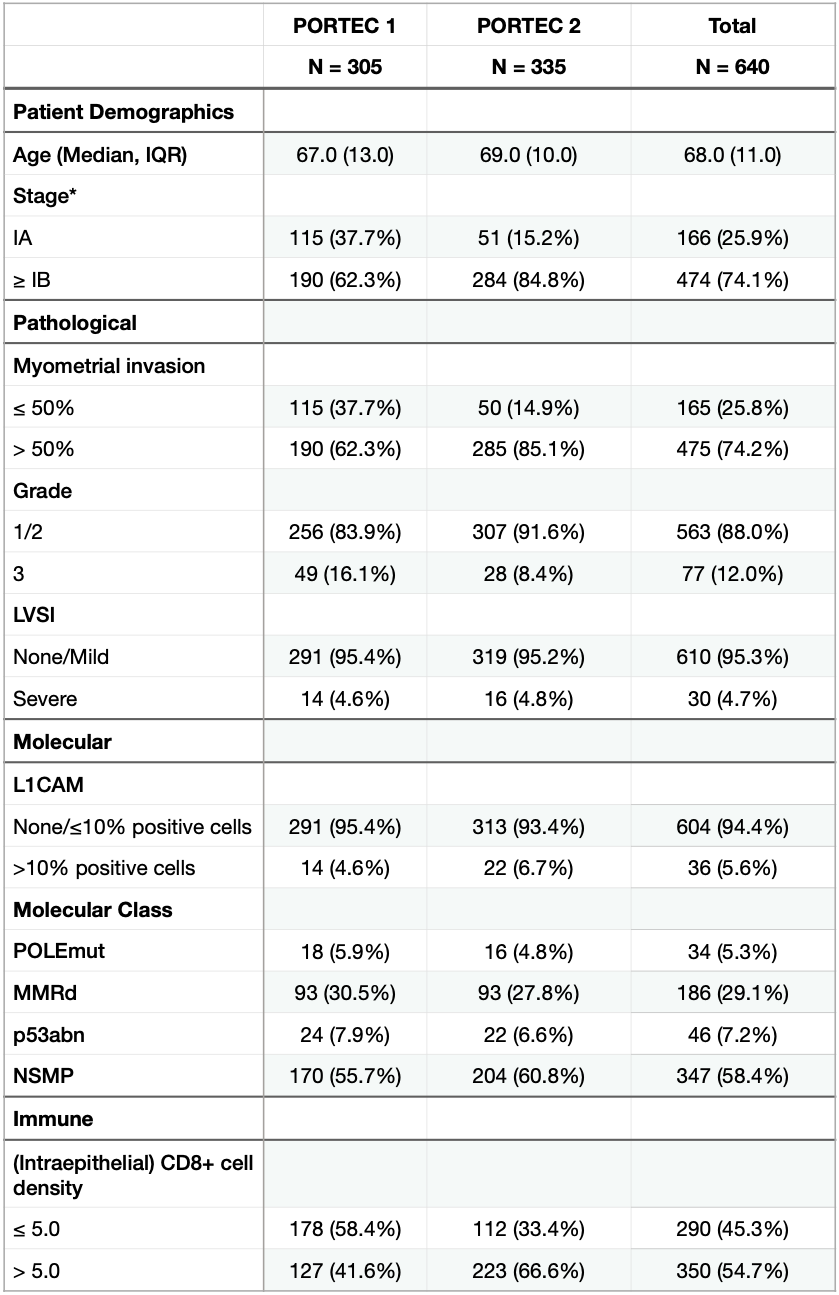} 
\caption{The characteristics of study participants for PORTEC 1 and 2. * After a posteriori central review 2 cases were classified as stage II, 2 as stage IIIA and 1 as stage IIIB.} 
\label{supp_tab3} 
\end{table}

\begin{table}
\centering
\includegraphics[width=0.8\textwidth]{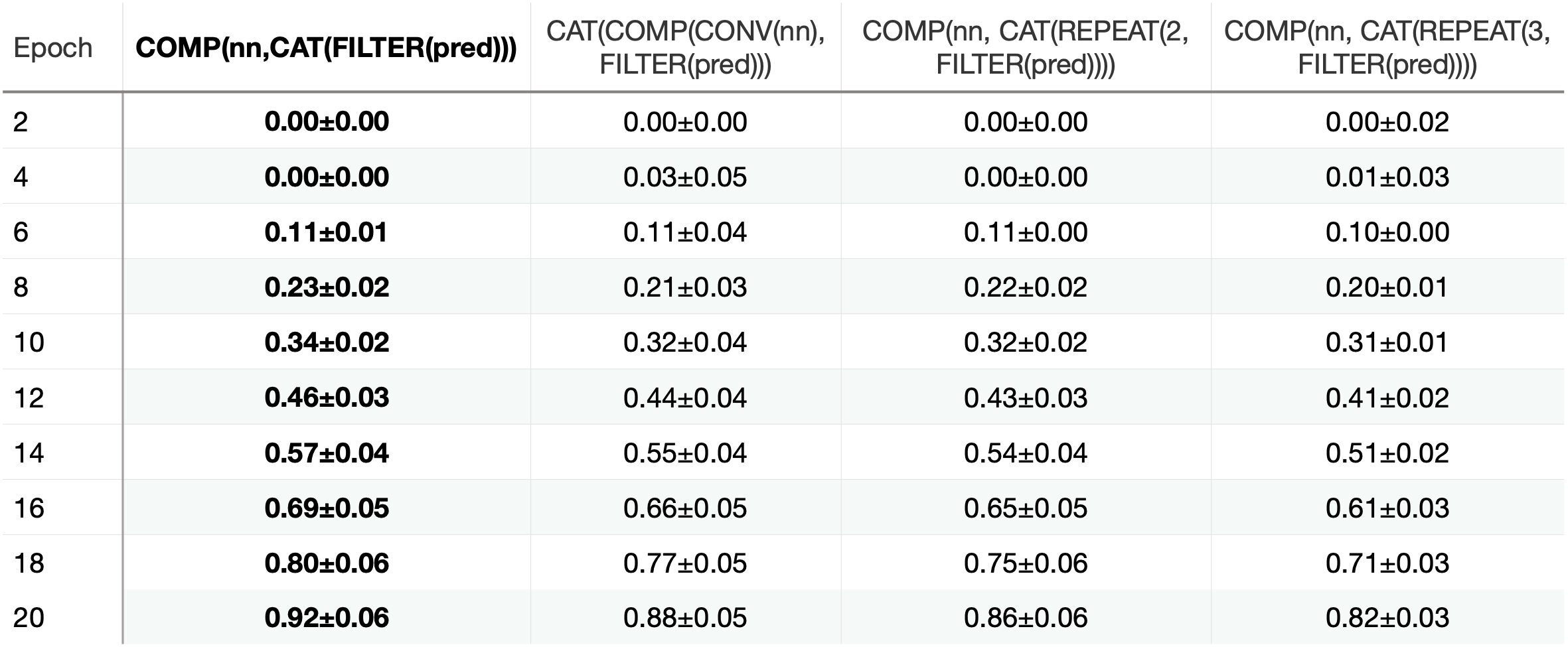} 
\caption{The JS score and its standard deviation of compared type-safe candidates for PORTEC study (PMI).} 
\label{supp_tab4} 
\end{table}

\begin{table}
\center
\begin{tabular}{l|llll}
\hline
                                                & Task         & Typed  & Functional  & Code availability \\ \hline
NTPT\cite{gaunt2017differentiable}           & Misc.        & \xmark      & \xmark & \xmark  \\
NS-CL\cite{mao2018neuro}                     & VQA          & \xmark      & \cmark & \cmark  \\
Prob-NMN\cite{vedantam2019probabilistic}     & VQA          & \xmark      & \cmark & \cmark  \\
DreamCoder\cite{ellis2020dreamcoder}         & Misc.        & \cmark      & \cmark & \xmark  \\
\textbf{HOUDINI}\cite{valkov2018houdini} & \textbf{Misc.} & \textbf{\cmark} & \textbf{\cmark} & \textbf{\cmark}
\end{tabular}
\caption{The comparison between existing program synthesis languages.}
\label{supp_tab5} 
\end{table}

\begin{table}
\centering
\includegraphics[width=0.8\textwidth]{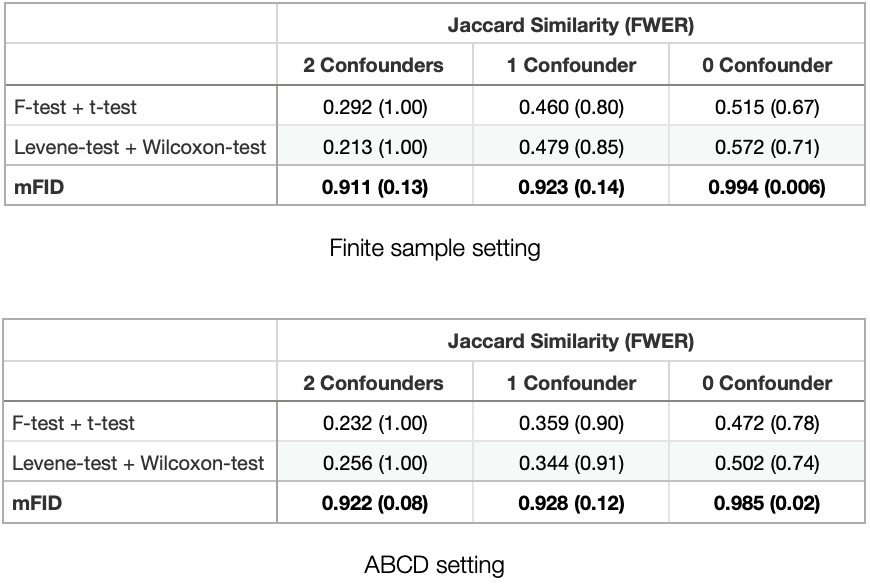} 
\caption{The comparison of statistical measurements for toy experiments.
Top: The results of the compared statistical measurements for the Finite sample setting. Bottom: The results of the compared statistical measurements for the ABCD setting. Here, all the measurements are applied for training the same type-safe function $\mathsf{COMP}(\mathsf{nn}, \mathsf{CAT}(\mathsf{FILTER}(\mathsf{pred})))$ under the proposed causal differentiable learning scheme. F-test + t-test is used in ICP and AICP. Levene-test + Wilcoxon-test is used in NICP.} 
\label{supp_tab6} 
\end{table}

\begin{table}
\centering
\includegraphics[width=0.8\textwidth]{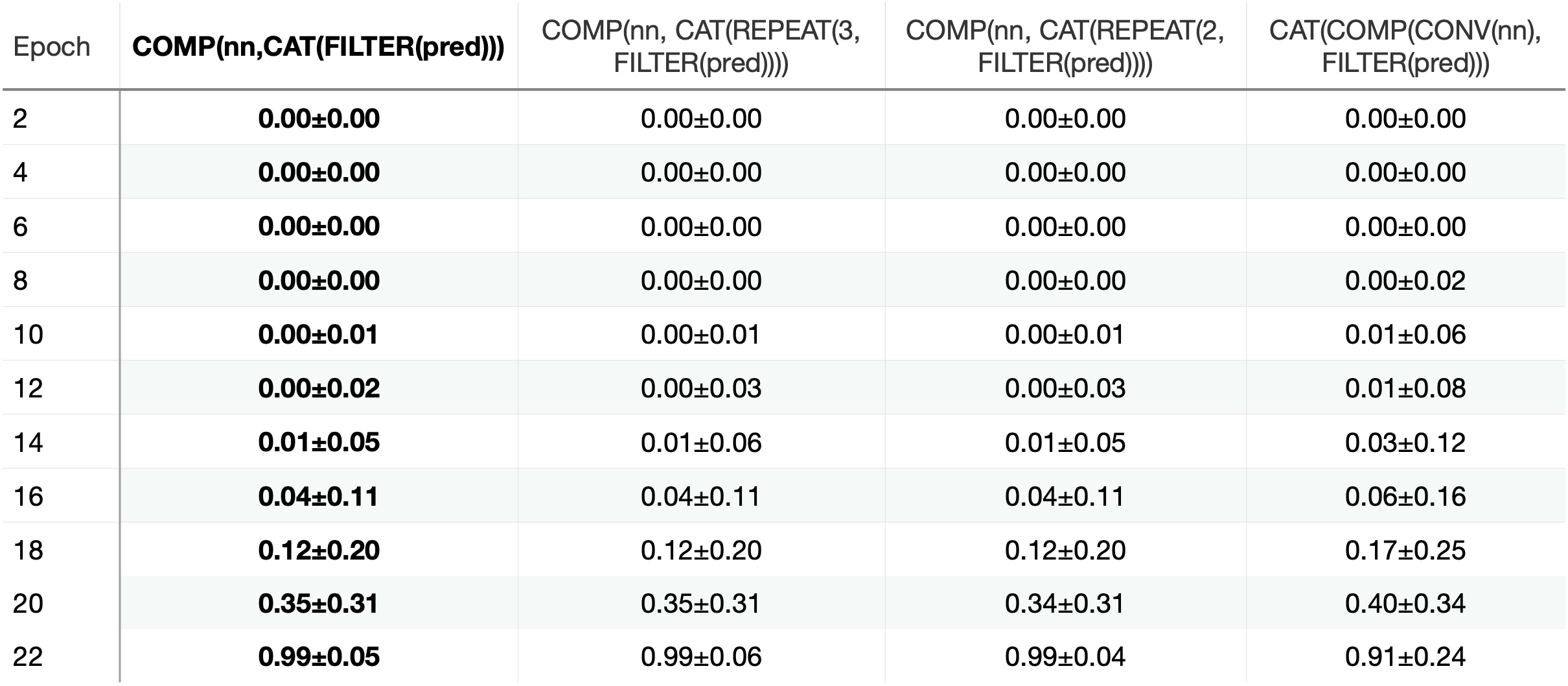} 
\caption{The JS score and its standard deviation of compared type-safe candidates for toy experiments (Finite sample setting).} 
\label{supp_tab7} 
\end{table} 

\begin{table}
\centering
\includegraphics[width=0.8\textwidth]{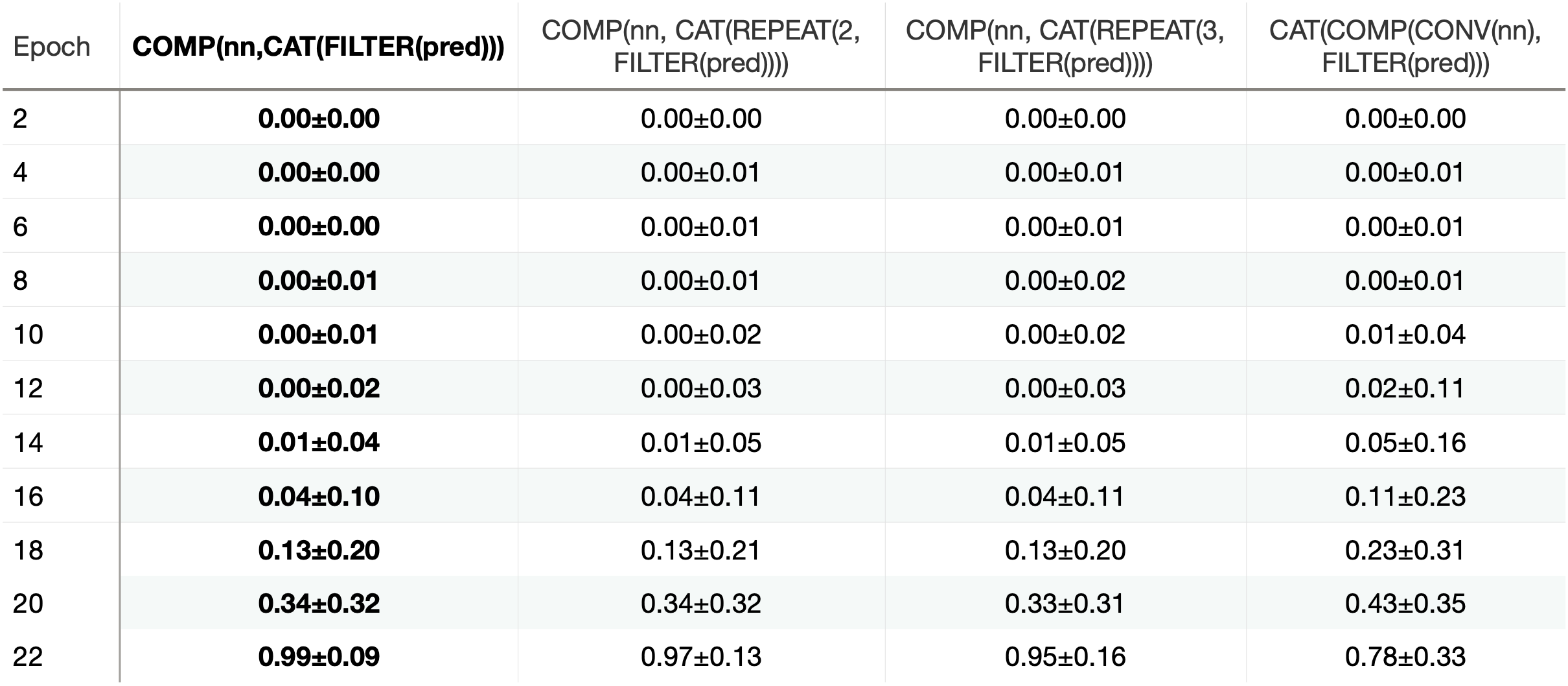} 
\caption{The JS score and its standard deviation of compared type-safe candidates for toy experiments (ABCD setting).} 
\label{supp_tab8} 
\end{table} 

\end{document}